# Crystal Growth, Terahertz Generation and Optical Characterization of Sodium Mesitylene Sulphonate (SMS)Crystal

Yamuna Murtunge,[†,‡] Vidhyadhar Patil,[¶] Ruturaj Puranik,[†] Jayakrishnan S S,[§] D Bansal,[§] Arijit Maity,[∥] Ravindra Venkatramani,[∥] S.B. Kulkarni,[‡] A Thamizhavel,[†] and S.S. Prabhu[*,†]

[†]Department of Condensed Matter Physics and Material Science, Tata Institute of Fundamental Research, Mumbai-400005

[‡]Institute of Science, Dr Homi Bhabha State University, Mumbai-400032

[¶]Department of Chemistry, Indian Institute of Technology Bombay, Mumbai-400076

[§]Department of Mechanical Engineering, Indian Institute of Technology Bombay, Mumbai-400076

[∥]Department of Chemical Science, Tata Institute of Fundamental Research, Mumbai-400005

E-mail: murtungeyamuna31@gmail.com, shriganesh.prabhu@gmail.com

## Abstract

An optically high-quality single crystal of sodium mesitylene sulfonate crystal was successfully grown by a slow evaporation method using methanol as solvent at room temperature. Single-crystal XRD has characterised the material and belongs to a monoclinic structure with a C2 space group. The lattice parameters are a=8.6926Å, b=7.3679 Å, c= 16.4519 Å, and α= γ =90°, β=103.7790° with unit cell volume of 1023.36 Å$^3$. Functional groups were determined using Fourier-transformed infrared spectroscopy. The optical quality of the generated crystal was evaluated using UV-VisNIR spectral analysis, which is transparent in the range of 300-1500 nm. We report the optical properties using terahertz time-domain spectroscopy (THz-TDS) and THz generation using crystal.

## 1. Introduction

In the last decade, generating few-cycle, high-peak electric field terahertz (THz) radiation has led [1–4], to various scientific applications [5] and exploration of different avenues of strong light-matter interactions [6]. A conventional technique of generating short THz pulses is by exciting a biased photoconductive switch with a femtosecond laser with a high repetition rate of the order of MHz While this method enables high optical-to-THz efficiencies [7], it is not scalable to large pulse energies due to the saturation of the THz electric field amplitude [8]. Hence, electro-optic crystals have been extensively researched [9] for generating and detecting THz radiation in the past few decades.

We present the synthesis and growth of sodium mesitylene sulphonate crystals. This crystal has been subjected to single crystal XRD, NMR, FTIR, UV-Vis-NIR, THz TDS and THz generation. We have also discussed the molecular structure and simulated the IR spectra.

## 2. Experimental Section

### 2.1 Synthesis

Sodium Mesitylene Sulfonate is prepared using the reaction shown in Figure 1. NaOH was added to a solution of 15 ml of methanol in a 100 ml R. B. flask (0.9152 gm, 22.88 mmol. 1 equiv.) under continuous stirring to form a solution. This solution was then cooled externally with wet ice.

2,4,6 trimethylbenzene sulfonyl chloride (5 gm, 22.88 mmol., 1 equiv.) was introduced slowly to the above reaction mixture; white-coloured precipitation of NaCl was observed during this process [10]. After completion of the reaction (during which reaction pH was 7), the whole reaction mass was filtered to remove precipitates. The filtrate was evaporated over a rota-vapor to remove excess methanol. The obtained semisolid mass was dissolved in 50 ml of dichloromethane to form a solution. This solution was washed with 50 ml water, dried over sodium sulphate, and concentrated. 1H NMR (800 MHz, DMSO-d6) δ 6.75 (s, 2H), 3.34 (s, 3H), 2.50 (s, 6H), 2.18(s,3H).

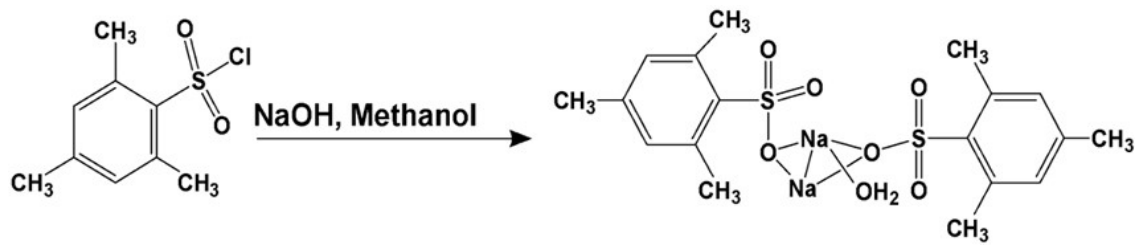

Figure 1: Chemical reaction depicting formation of the crystal.

## 2.2 Crystal Growth

Sodium Mesitylene Sulfonate (SMS) crystal was obtained from a supersaturated SMS solution by slow evaporation [11] at ambient temperature. The saturated SMS solution was prepared with methanol as a solvent, filtered, transferred to a round bottom flask, and covered with a polythene sheet. After 15 days, we made a few holes in the polythene sheet for slow evaporation. After 30 days, we obtained SMS single crystal with 10 × 5 × 0.182 mm3 and 10 × 5 × 1.266 mm3 dimensions [12]. The grown crystals were well-shaped, transparent, colourless, and chemically stable under atmospheric conditions. These are shown in Figure 2 and are free of visible defects.

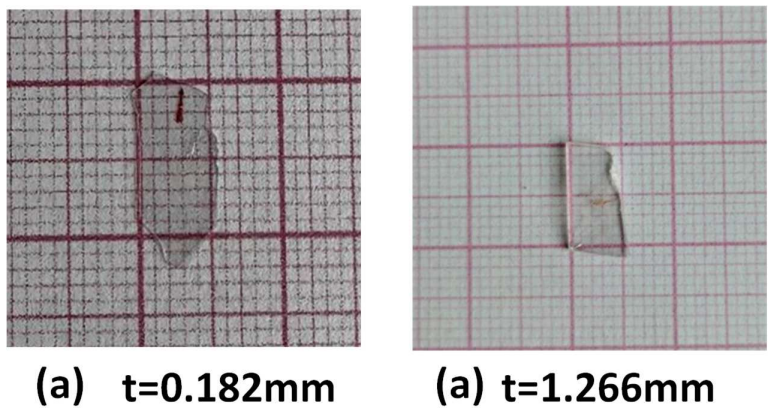

(a) t=0.182mm    (a) t=1.266mm

Figure 2: Photographs of the crystals grown are shown with their thickness (a)0.182 mm and (b)1.266 mm used for the experiments

## 2.3 Computational Details

Electronic calculations were performed in the DFT framework in Vienna Ab initio Simulation Package (VASP) [13–15] on the primitive cell with 53 atoms with plane wave cutoff energy of 800eV for convergence. The number of points in the Brillouin zone for self-consistent DFT calculation is 6x6x2. We used a generalized gradient approximation (GGA) in the Purdew-Burke-Ernzerhof (PBE) parametrization for solids [16]. Atom positions were relaxed till forces in the system were less than 1 meV/A. The optimized structure is shown in Figure 6(a). Gamma point phonons were found using the density functional perturbation theory (DFPT) framework implemented in the VASP. This is shown in Figure 1(b). Harmonic vibrational frequencies were calculated using the optimized structure of the SMS molecule in both vacuum and methanol (PCM). Density functional theory was deployed to calculate the electronic structure where CAM-B3LYP exchange-correlation functional and 6-311++G(d,p) [17] basis set were used. All the single molecule calculations were performed in the Gaussian09 [18] software package, and Gauss view 5.0.8 was utilised for visualisation and frequency mode analysis.

## 3 Results and discussion

### 3.1 Single crystal XRD And Laue Diffraction:

The grown crystal belongs to the Monoclinic structure with the non-centrosymmetric C2 space group and Z=2. The lattice parameters of the crystals a=8.6926Å, b=7.3679Å, 4 c=16.4519Å, and α=γ=90°, β=103.7790°, were determined using Bruker D8 Venture single crystal X-ray diffractometer with CuKα(λ=1.54178Å) radiation at room temperature. The hydrogen bond[O(4)-H(4)...O(2)] is formed between the hydroxyl group acting as cation and sulfonate moiety as an anion with a bond distance and a bond angle of 2.841Å and 147°, respectively, which stabilised the molecule.

Table 1: Hydrogen bonding interaction in SMS crystal

| D-H...A | d(D-H) | d(H....A) | d(D...A) | (DHA) |
|---|---|---|---|---|
| O(4)-H(4)...O(2) | 0.84 | 2.10 | 2.841 | 147 |

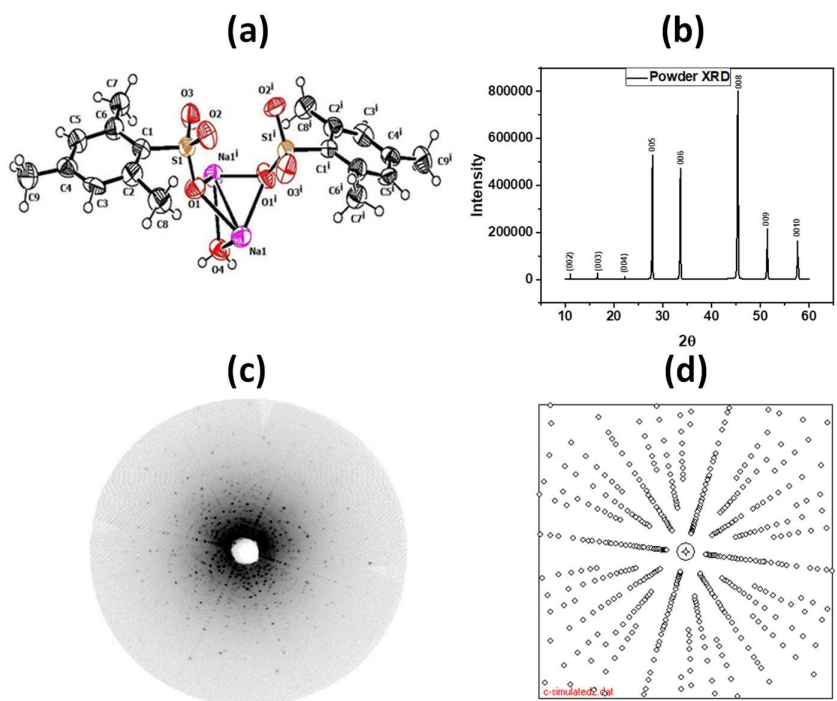

Figure 3: The Chemical formula of (a)ORTEP-II at 50% probability is shown. (b)XRD pattern for the single crystal(c)Photograph of the Laue pattern and (d)Simulated Laue pattern of the crystal(001).

Table 2: Crystal data and structure refinement for SMS

| Empirical formula | $C_{18}H_{24}Na_2O_7S_2$ |
|---|---|
| Formula weight | 462.47 |
| Temperature | 297(2) K |
| Crystal system | Monoclinic |
| Space group | C2 |
| Unit cell dimensions | a = 8.6926Å, α = 90° |
| | b = 7.3679Å, β = 103.7790° |
| | c = 16.4519Å, γ = 90° |
| Volume | 1023.36(4)Å$^3$ |
| z | 2 |
| Density (Calculated) | 1.501Mg/m3 |
| Absorption coefficient | 3.123mm−1 |
| F (000) | 484 |
| Crystal size | 0.078 × 0.067 × 0.023 mm3 |
| Theta range for data collection | 2.765°to 70.093° |
| Index ranges | −10 ≤ h ≤ 10, −8 ≤ K ≤ 8, −20 ≤ l ≤ 20 |
| completeness to theta | 67.67 to 99.9 |
| Refinement method | Full matrix least squares on F 2 |
| Data/restraints/parameters | 1916/2/178 |
| Goodness-of-fit on F2 | 1.076 |
| Final R indices [I > 2σ(I)] | R1 = 0.0537, wR2 = 0.1407 |
| R indices (all data) | R1 = 0.0635, wR2 = 0.1520 |
| Absolute structure parameter | 0.03 |
| Extinction coefficient | 0.0092 |
| Largest diff. peak and hole | 0.387 |

The XRD of a single SMS crystal shows a prominent intensity peak along the lattice parameters of c(001), indicating the crystalline nature of the grown single crystal and the crystal's orientation along the c direction shown in Figure 3(b). Laue's back-reflected simulated diffraction patterns along (001) are well-matched with the experimental Laue diffraction images shown in Figures 3(c) and 3(d). The data from XRD provide valuable insights into the crystallographic structure of sodium mesitylene sulfonate crystal. The ORTEP-II at 50% probability is shown in Figure 3(a).

## 3.2 FTIR:

The sample was subjected to Fourier infrared spectroscopy (FT-IR-JASCO) to identify the vibrational modes and determine the functional groups. We used a 1 mm thick pellet prepared with a 1:9 KBr and SMS powder ratio for characterisation. The spectral data were collected in the region of 400 cm−1–4000 cm−1 . The spectral analysis was carried out based on the vibrations of the mesitylene sulfonate molecule. Figure 4(a) shows that several peaks were detected, informing the complex structure of a material. The peaks at 539 cm−1 and 1458 cm−1 are assigned to C=C stretching vibrations indicating unsaturated compounds or aromatic rings [19],and the peak at 1600 cm−1 is due to the C=C stretching vibration of the molecule. A narrow band at 983 cm−1 is due to the C-H out-of-plane bending vibrations. The peak at 2949 cm−1 , describing simple carbonyl compounds, results from the bonding between methyl (CH3) and the benzene ring. The peak at 1035 and 1351 cm−1 represents the existence of a sulfonate group in the structure. The strong peak around 1174cm−1 is due to S=O stretching vibration in the molecule.

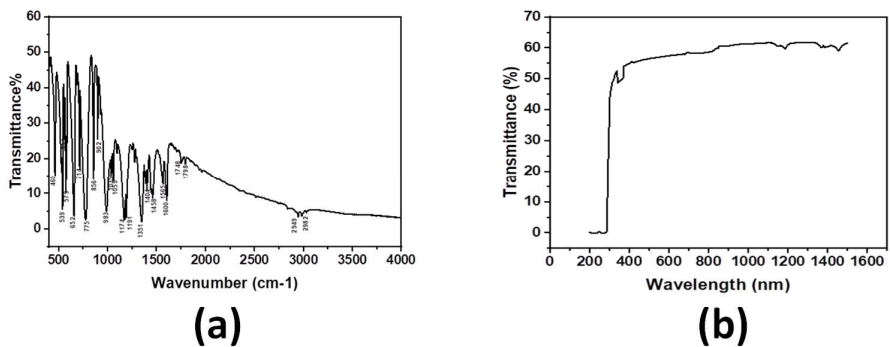

Figure 4: (a) FTIR spectrum of the crystal and (b) UV-VIS-NIR spectrum of the crystal are shown at room temperature

### 3.3 UV-VIS-NIR Spectroscopy:

The optical transmission spectrum of the titled crystal was recorded in the region of 250–1500 nm using a JASCO UV-Vis-NIR spectrometer with a thickness of 0.182 nm. This spectrum is shown in Figure 4(b). It is observed that the lower cut-off of this crystal is at 285 nm [20]. The sample may also be transparent below 200 nm, but this data is insufficient to conclude due to the spectrometer sensitivity below 285 nm. The crystal is found to be transparent in the entire region of the visible and near IR spectrum, suggesting this crystal is suitable for the generation of second harmonic devices [21] as well as for optoelectronic devices.

### 3.4 NMR:

The 1H NMR spectrum of Sodium Mesitylene Sulfonate was recorded using DMSO d6 as a solvent on a Bruker Biospin 800 MHz and measured using a 5 mm triple resonance probe. The number of hydrogens in mesitylene sulfonate molecules [22] matches hydrogen peaks at 1H NMR. The NMR spectra are shown in Figure 5.

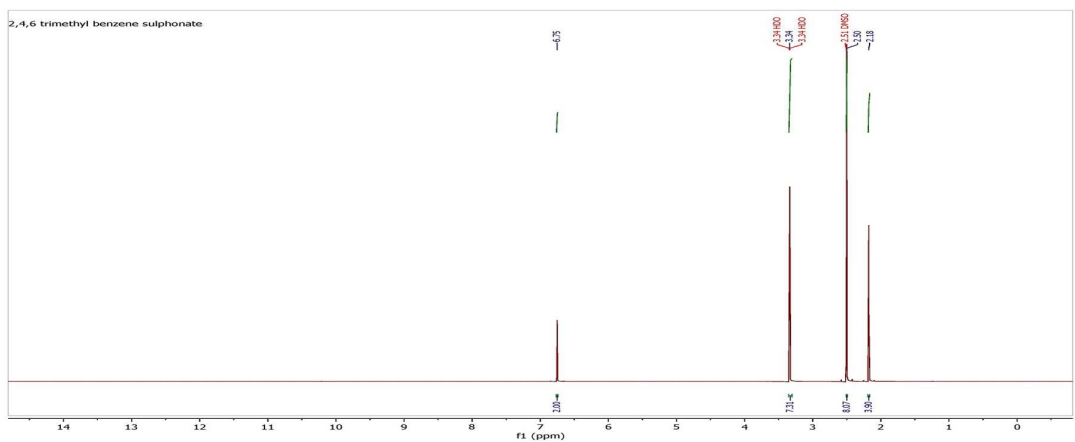

## 3.5 Computational Details:

We performed the optimization and frequency calculation in different conditions, like with/without explicit water molecules and with/without Na ions in the gas phase (SI-4). After comparing the data with the experiment, Na ion and water show significant IR spectra, so we decided to go with explicit water and Na ion. Then, we performed the same calculation in methanol (Polarizable Continuum Model) to check the effects of the solvent on the optimized structure. We compared the high-frequency vacuum and methanol (PCM) modes for the Na ion and water molecule combination. The high frequencies of the solvent (PCM) structure almost match the solid IR shown in (SI-3). The absorption peak at 0.64 THz is attributed to the rocking motion of the -CH3 molecule attached to the benzene ring. The peaks at 0.945 THz and 1.92 THz are due to the vibrational in-plane bending of water and methyl moieties. We found that the simulated IR spectrum in solid state DFT (VASP) exhibits a frequency shift of 0.299 THz for 0.9498 THz and 0.2098 THz for 1.922 THz towards higher frequencies than the experimental results. In contrast, for the single molecule (PCM), the IR vibrational frequencies have shifted to a lower frequency. As we performed the DFT(PCM) calculation on a single molecule, the obtained results showed a significant deviation from the experimentally measured frequencies.

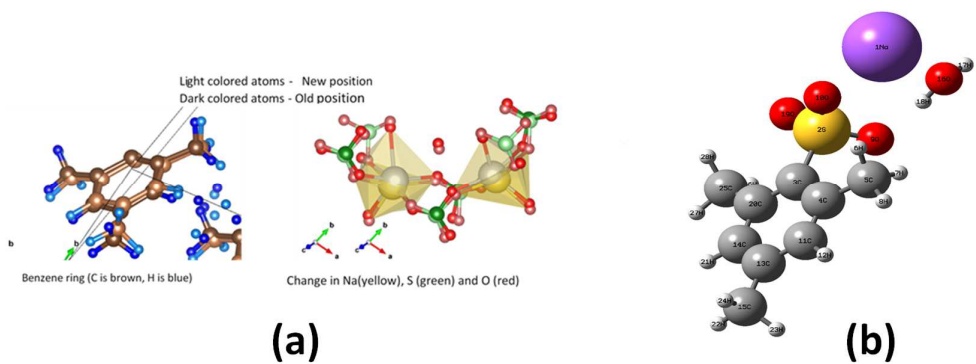

(a) (b)

Figure 6: The optimized molecular structures using(a) solid-state DFT and (b)Single molecule with PCM model calculated at CAM-B3LYP/6-31++g(d,p)

Polar electron donor and acceptor groups within the molecule create an asymmetric charge distribution. This structural asymmetry enhances the polarizability of the molecule and consequently enhances the first hyperpolarizability The total molecular Dipole moment, average polarizability and first hyper-polarizability were calculated using the following equations: [20, 23]

$$\mu = (\mu_x^2 + \mu_y^2 + \mu_z^2)^{1/2} \quad (1)$$

$$\alpha = (\alpha_{xx} + \alpha_{yy} + \alpha_{zz})/3 \quad (2)$$

$$\beta = (\beta_x^2 + \beta_y^2 + \beta_z^2)^{1/2} \quad (3)$$

The calculated electronic dipole moment, average polarizability and first hyper-polarizability of the optimised geometry are 10.5491D, -11.75×10−24, 2.723×10−30 esu respectively, compared to the Urea [24] total dipole moment, polarizability and first hyperpolarizability are 8, 2.81, and 7 times greater in the SMS single molecule. The level of theory and basis set for both Urea and SMS are different. As we are dealing with anion, we included diffused basis set (++) in the SMS molecule; however, urea is a neutral molecule. The molecule exhibits a significant first hyperpolarizability along the βxxx direction shown in 7(b), indicating dipole moments are oriented in this direction [25]. The total first-order hyperpolarizability value of the SMS molecule indicates a good candidate for a non-linear optical material. This comparison between theory and experiment establishes the reliability and applicability of our computational approach to predicting molecular and vibrational properties.

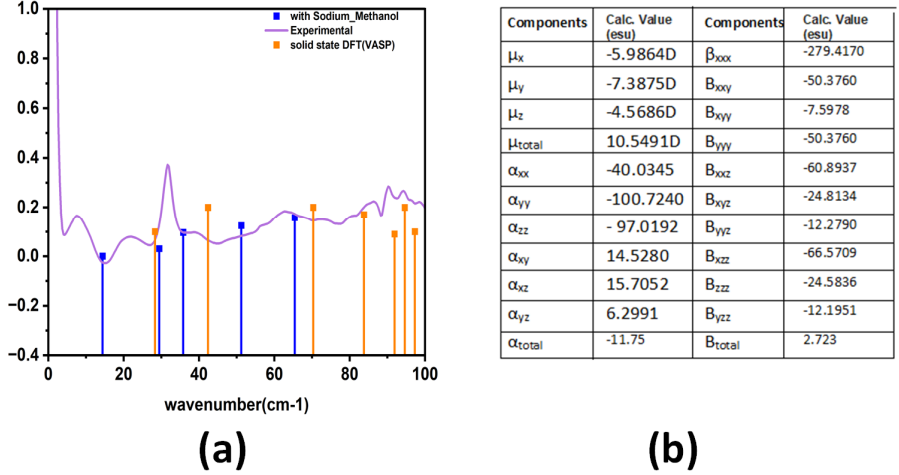

Figure 7: (a)the simulated Infra-Red spectrum using solid-state DFT and PCM model with methanol as solvent are shown with the measured one (b)The dipole moment(μ), polarizability(α), and first-order hyperpolarizability(βtot) values of SMS molecule

## 3.6 Tera-Hertz time-domain spectroscopy (THz-TDS):

We conducted THz transmission measurements of the SMS crystal normal along the < 110 > plane. The THz transmission of the sodium mesitylene sulfonate crystal is assessed using the experimental setup described above. This setup (frequency resolution of 4 GHz) incorporates a Femtosecond pulsed laser, FemtoSource Synergy (Femtolasers GmbH), generating 10 fs pulses at an 80 MHz repetition rate with a centre wavelength of 800nm. The setup includes four parabolic mirror configurations, as shown in Figure 8 [26]. The laser beam is divided into the pump and the probe beam. The pump beam is utilised to generate THz radiation employing an LT-GaAs-based photo-conductive antenna (PCA), while the probe beam is responsible for detecting the THz signal. Using a standard electro-optic method, the resulting THz signal is observed using a 2 mm thick ZnTe crystal with < 110 > orientation.The entire experimental arrangement was contained within a nitrogen-purged box to prevent any unwanted absorption of water lines in the air. We obtain a dielectric constant and conductivity using a complex refractive index with thick crystal approximation.

$$\tilde{n} = n - i\kappa \tag{4}$$

$$n = \left(\frac{c\varphi(\omega)}{\omega d}\right) + 1 \tag{5}$$

$$\kappa = \frac{c}{\omega d} \cdot \ln\left(\frac{4n}{\rho(\omega)(n+1)^2}\right) \tag{6}$$

where, the speed of light in a vacuum is c, the thickness of the sample is d, ω is the frequency, ϕ(ω) is the phase difference, and ρ is the transmittance. At the THz frequency range, we observed a strong interaction between the vibrational modes of the crystal and THz electromagnetic radiation. This interaction results in the absorption and scattering of THz radiation by phonons, leading to changes in the refractive index. In Figure 9(a), we present the extracted real component of the refractive index (black curve) in the THz frequency region. It shows a resonance feature around 1 THz, and the refractive index varies quite a bit compared to the refractive index measured in the optical (red curve) wavelength domain (300 nm to 1000 nm). It varies from ~1.482 (300nm) to ~1.469 (1000nm), which is quite small compared to the THz range, and we do not see any resonance features in this optical range.

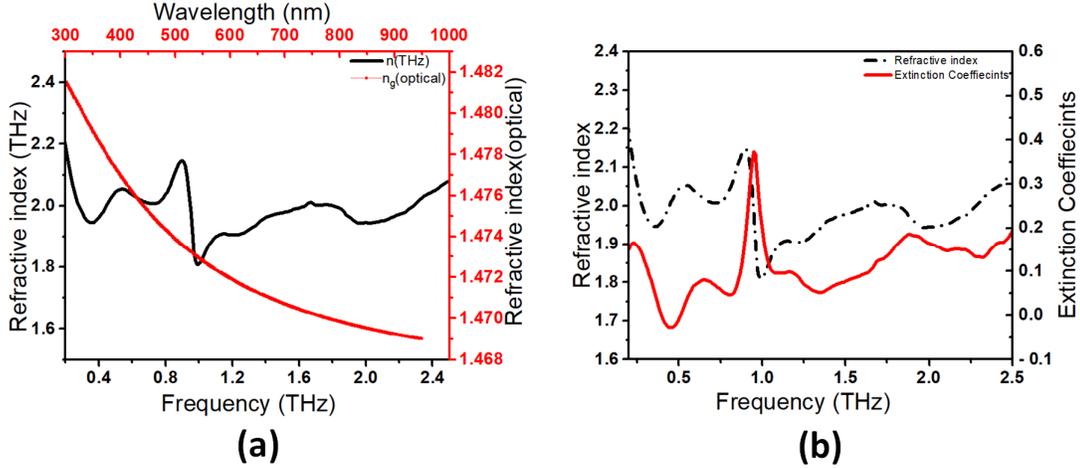

(a)                                         (b)

In Figure 9(b), both the real (black curve) and imaginary (red curve) parts of the refractive index are shown, and the imaginary part clearly shows a Lorentzian resonance peak around 1 THz. The resonance peaks obtained are related to the oscillatory modes in the crystals and are responsible for the specific absorption modes. All observed modes are IR active. Due to the incident electromagnetic radiation, these can resonate at a particular frequency. Additionally, it causes an increase in the induced polarisation of the material, contributing to changes in the dielectric constant. The experimentally determined real and imaginary parts of the dielectric constants were fitted using the Lorentz multiple oscillator model [27] given as:

$$\tilde{\varepsilon} = \varepsilon_\infty + \sum_{m=1}^{n} \frac{g_m}{(\omega_m^2 - \omega^2) + i\omega\gamma_m} = \varepsilon'(\omega) + i\varepsilon''(\omega) \quad (7)$$

where $\varepsilon_\infty$ is the high-frequency dielectric constant, $g_m$ (= $\Delta\varepsilon_m \omega_m^2$) represents the oscillator strength for the m$^{th}$ resonance, $\Delta\varepsilon_m$ is the dielectric strength, $\omega_m$ is the resonance frequency, and $\gamma_m$ is the damping factor. The theoretical fits of both real ε ′ (ω) and imaginary ε ″ (ω) dielectric constant curves shown in Figure 10 (b) exhibit good agreements with experimental THz spectra tabulated in Table 4. The deviations in the fitted theoretical spectra are well within the range of the experimental errors.

| Oscillator m | Resonant frequency $\omega_m$ (THz) | Oscillator strength $g_m$ | Scattering Time τ (ps) |
|---|---|---|---|
| 1 | 0.2295 | 0.0203±0.0061 | 10.04199±3.07 |
| 2 | 0.6525 | 0.0041±0.0066 | 41.5552±1.8 |
| 3 | 0.940 | 0.1154±0.0097 | 17.2376±1.84 |
| 4 | 1.922 | 0.1086±0.0337 | 7.74879±2.87 |

As can be seen from the fitted parameters, there Table 4: Dielectric properties of SMS crystal extracted from Lorentz Oscillator fitting Model are several minor resonances with two primary modes. The strongest of these is at ~0.94 THz with a scattering time of ~17 ps followed by a broad one at ~1.9 THz with a scattering time of ~7 ps. The observed resonance frequencies are pretty close to the calculated ones by DFT. The deviation can be partially attributed to temperature as DFT vibrational mode frequencies are at 0 K, while the measured values are at 300 K. The oscillator strengths of these modes are also comparable. The other two weak modes show that the crystal has lowfrequency vibrations. However, they do not contribute significantly to the dielectric response. Furthermore, our study of complex conductivity on the same crystal demonstrated that the real conductivity increases at these resonant frequency modes.

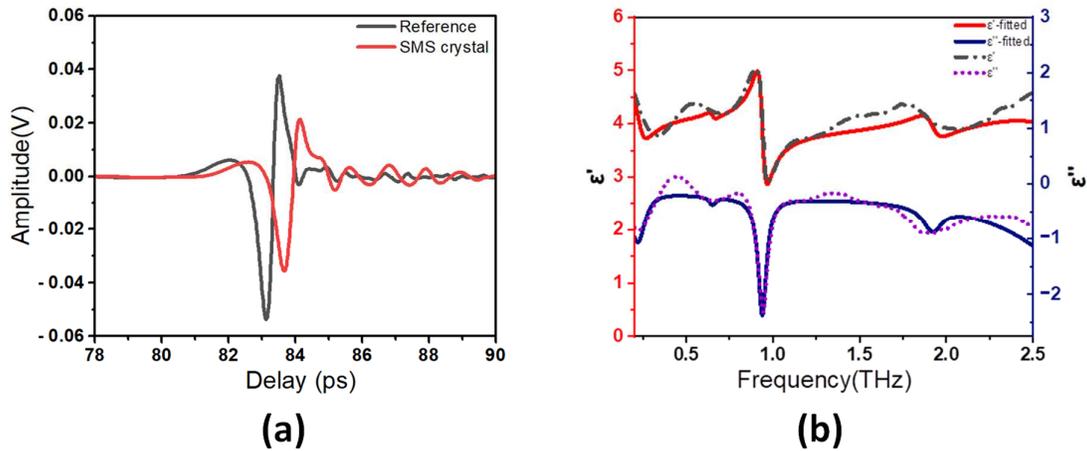

(a)          (b)

Figure 10: (a)THz time domain transmission pulse through SMS crystal (red) is shown with the reference (black) pulse. (b) Experimentally obtained real $\varepsilon'$ and imaginary $\varepsilon''$ are shown with fitted equations (see text)

The absorption seen in THz radiation at those frequencies may be attributed to the free charge carriers being excited or promoting existing charge carriers to higher energy and mobility states, consequently enhancing the conductivity at those frequencies.

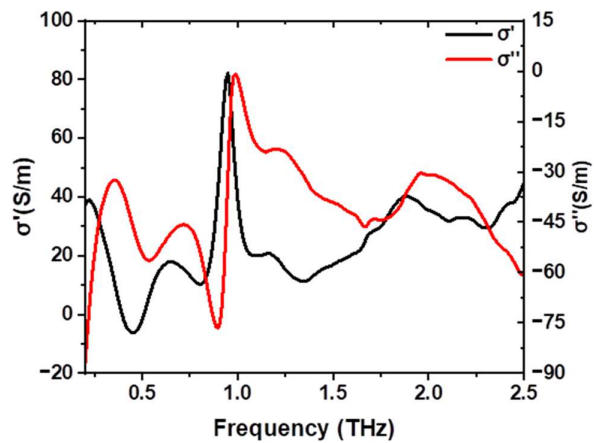

Such crystals can then be used as an effective filter at those THz frequencies where the transmission can be attenuated.

### 3.7 THz Generation

An as-grown 0.182mm thick SMS crystal was employed without polishing and cutting through optical rectification using an amplified laser with a centre wavelength of 800nm and a repetition rate of 1 KHz to generate the THz radiation. The pulse duration, single pulse energy and pump beam diameter were approximately 45 fs, 5mJ and 11mm, respectively. The generated THz pulse with a good signal-to-noise ratio was detected using an electro-optic sampling in a < 110 > oriented 1 mm thick ZnTe crystal. The crystal shows small dimples around 1.04THz and 1.72THz. The scan was taken at a 140-degree angle rotation. Unlike other nonlinear crystals in the THz domain, THz generation from this crystal is inefficient or could be due to the limitation of the ZnTe detector.

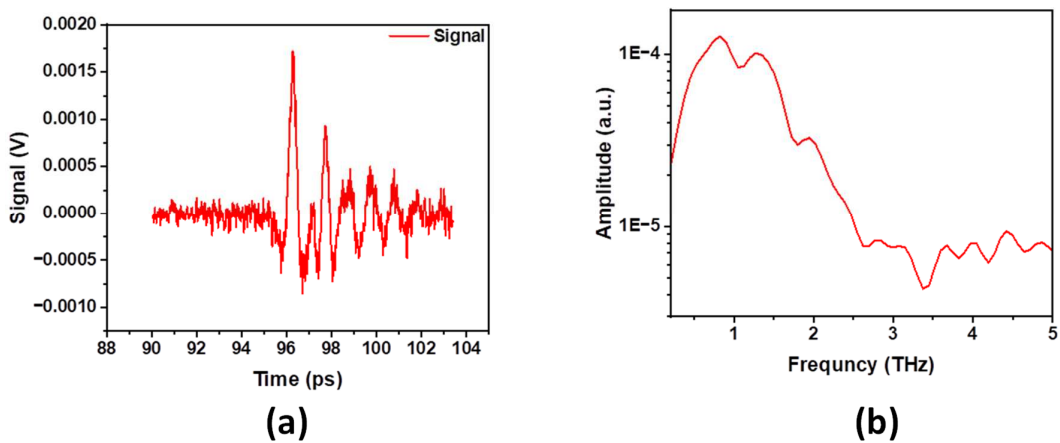

Figure 12: a) THz transient electric field and (b)Fourier transformed THz spectrum generated in 0.182mm thick SMS crystal

### 4. Conclusion

The nonlinear optical organic SMS crystal was synthesized and grown using a slow evaporation technique. Single crystal XRD was utilized to ascertain the lattice parameters and crystal structure. Confirmation of functional groups and molecular structures was achieved through FTIR and NMR spectral analysis. The electronic and molecular characteristics were investigated using the Density Functional theory framework. The complex refractive index, dielectric constant, and conductivity have been characterised using THz time-domain spectroscopy in the 0.2–2.5 THz region. The SMS crystal generates THz spectra from 0.2 to 2.5THz with few dimples at 1.04THz and 1.72THz. Since this crystal is non-centrosymmetric, work is underway on polarization-dependent time-domain spectroscopy.

### Acknowledgement

The authors sincerely thank the Tata Institute of Fundamental Research (TIFR) Mumbai for providing the necessary research resources. We acknowledge IIT Madras(SAIF)for conducting the single crystal XRD analysis. The authors thank Mr Gajendra Mulay, Mr Balasaheb Chandanshive, Dr Shraddha Chaudhary, Dr Malay Patra, Dr Shreyas, and Ms. Ruta Kulkarni for their invaluable contributions and support throughout this research.